# Determination of the drop size during atomization of liquid jets in cross flows

T.-W. Lee, Jung Eun Park
School of Engineering for Matter, Transport and Energy
Arizona State University
Tempe, AZ 85287-6106

and

Ryoichi Kurose
Department of Mechanical Engineering
Kyoto University
Kyoto, Japan

Corresponding author: T.-W. Lee, attwl@asu.edu

Department of Mechanical and Aerospace Engineering
Arizona State University
Tempe, AZ 85287, USA

**ABSRACT-** We have used the integral form of the conservation equations, to find a cubic formula for the drop size in liquid sprays in cross flows. Similar to our work on axial liquid sprays, the energy balance dictates that the initial kinetic energy of the gas and injected liquid be distributed into the final surface tension energy, kinetic energy of the gas and droplets, and viscous dissipation incurred. Kinetic energy of the cross flow is added to the energy balance. Then, only the viscous dissipation term needs to be phenomenologically modelled. The mass and energy balance for the spray flows renders to an expression that relates the drop size to all of the relevant parameters, including the gas- and liquid-phase velocities. The results agree well with experimental data and correlations for the drop size. The solution also provides for drop size-velocity cross-correlation, leading to drop size distributions based on the gas-phase velocity distribution. These aspects can be used in estimating the drop size for practical applications, and also in computational simulations of liquid injection in cross flows.

**Nomenclature**

$A_c$ = cross-sectional area of the spray

$A_{inj}$ = injector exit area

$d_{inj}$ = injector diameter

$D$ = drop diameter

$D_i$ = drop diameter for the i-th size bin

$D_{32}$ = SMD = Sauter mean diameter

K, K'= proportionality constants for the viscous dissipation term

n = drop number density

p(D) = normalized drop size distribution function

q = momentum ratio = $(\rho_L u_{inj}^2)/(\rho_g u_{in}^2)$

$u_{in}$ = velocity of the incoming gas

$u_{inj}$ = mean injection velocity

$u_L$ = mean drop velocity

$u_{out}$ = velocity of the outgoing gas

$\left\langle \left( \dfrac{\partial u}{\partial y} \right) \right\rangle$ = average velocity gradient in the spray

V = volume of the spray bounded by A and spray length

$\mu_L$ = liquid viscosity

$\rho_g$ = ambient gas density

$\rho_L$ = liquid density

$\sigma$ = surface tension

# INTRODUCTION

Liquid jets in cross flows are of interest in combustion and other spray devices, and much work exists in experimental studies to determine the structure, drop size and velocity distribution in such flows [1-6]. In this flow geometry, the gas momentum (and energy) is the main cause for the disintegration and break-up of the injected liquid. This is in contrast to the straight pressure-atomized sprays where it is the liquid momentum (kinetic energy) that is the dominant driver of the subsequent atomization process. Liquid streaming into cross flows has important applications in gas-turbine engine combustion such as afterburners, low $NO_x$ burners, high-speed combustors (e.g. supersonic combustors), and also in cooling sprays for turbine blades and in rocket engine combustion. Recent works on liquid atomization in cross flows focus on gas-turbine applications [7-10]. As the fuel is injected in gas-turbine combustors in a swirl pattern at some angle with respect to high-speed incoming air, liquid jets in cross flows can be considered as a baseline geometry for more complex fuel injection in gas-turbine combustors.

The trajectories of the liquid as a function of the so-called momentum ratio ($q=\rho_L u_{inj}^2/\rho_g u_{in}^2$) are well characterized, and there appear to be reasonable agreements between different experimental results [1-3]. Faeth and co-workers [1] measure the drop size close to the curved liquid column, naming such drop formation process as the primary atomization. Subsequent "secondary" atomization then is still undetermined, at least not measured in their study [1]. Other studies present plots the Sauter mean diameter (SMD) as a function of the coordinates following the liquid trajectory, or at some arbitrary, fixed location downstream [1]. Unless the gas velocity is extremely high, the liquid and gas interact over a distance to produce droplets of varying size, which typically decreases as the distance from the injection point increases. The interaction of the gas and liquid continues, and the asymptotic final drop size is not always evident [1-3]. The "atomization length" in the gas streamwise direction will vary as a function of the flow conditions, and also a large test section will be required to capture this full atomization process. A few studies do present data for practical use toward gas-turbine applications, where the average drop size is measured for the entire plane downstream of the liquid column [4, 5, 8, 9]. Recent experimental data by Freitag and Hassa [4] and others [5, 6] are also useful and compared with the current theory later in this paper.

There are a few experimental correlations for the drop size [1, 4, 9, 12], but they tend to be limited in their scope. As noted above, Faeth and co-workers [1] only report the primary break-up drop size near the liquid core where the drop size increases as a function of the streamwise direction along the liquid column. A general correlation is presented in an earlier work by Ingebo [12], where the SMD is correlated with the product of Weber and Reynolds numbers. These correlations may be useful at high Weber numbers (large gas velocities), but they yield finite drop size even when the gas velocity is close to zero. Thus, a need exists for a theoretical basis for putting together the experimental results. Other structural data can also be found [13-17], where various effects on the fluid dynamic aspects of the liquid jets in cross flows are revealed. In particular, Eriksson [15] uses particle image velocimetry (PIV) and phase Doppler anemometry to determine the gas- and liquid-phase velocities, and the drop size. In spite of these sets of data, a theoretical formulation that encompasses all the relevant parameters would be quite useful in bringing the physical variables together in a fluid-dynamically sensible manner. Such a formulation can then be used across a wide range of conditions, based on available data.

Recently, we developed a theoretical framework using integral form of the conservation equations which led to a closed-form solution for the drop size in pressure-atomized sprays with and without swirl [**18**]. As will be described below, this method does not involve any ad-hoc assumptions or non-physical descriptions of the atomization process, and thus represents a new theoretical formulation for the spray atomization in cross flows. Lefebvre [19] had used a simple zero-dimensional relationship where the kinetic energy of the atomizing air is equated with the final surface tension energy. The rest of the energy terms and their effects are lumped into a parameter representing the efficiency of the atomizing process, so there are only two terms in this energy relationship and no viscous term. Therefore, as the title of that work ("Energy considerations in twin-fluid atomization [19]") indicates, it is a zero-dimensional energy consideration, not a full energy balance in a control volume setting. The current formulation is a complete set of mass, energy (and momentum), where the momentum balance has also been used in some of our previous work for a complete solution of the spray atomization problem [18, 20-22]. Current approach by-passes the complex physics of the atomization process by enveloping the spray in a control volume, where the incoming and outgoing energy terms are balanced according to the conservation of energy. The kinetic energy of the liquid and gas, in this case, must be distributed into the kinetic energy of the liquid and gas, surface tension energy and viscous dissipation. When combined with the conservation of mass, this method leads to a cubic formula for the drop size as a function of the injection parameters. This framework has worked quite well in predicting the drop size in pressure-atomized sprays with and without swirl, as a function of all the relevant injection parameters and liquid/gas properties [18, 20-22]. Some of the resulting expressions could also be cast in terms of Reynolds and Weber numbers, after some proper normalization [22]. Comparisons with experimental data show very good agreement, and also provide insights into the atomization process. In this work, we show that this approach can be extended to liquid jets in cross flows, simply by adding the kinetic energy balance for the gas crossflow. The formulation is based on fundamental conservation laws (of mass, energy, and momentum) cast in integral forms, and thus is an extension of this general formulation toward spray atomization in cross flows. Potential applications of this work include estimating the drop size in fuel or other liquid sprays in cross flows, and also in setting initial conditions for computational simulation of sprays.

**MATHEMATICAL FORMULATION**

The basic integral form of the conservation equations for mass, momentum and energy has been shown in our previous works [18, 20-22]. In particular, an expression for the viscous dissipation leads to the following closed-form solution for the drop size in pressure-atomized sprays [18].

$$D_{32} = \frac{3\sigma + \sqrt{9\sigma^2 + K' \frac{\rho_L \mu \bar{u}^2}{u_{inj}} \frac{u_{inj}^2 - \bar{u}^2}{2}}}{\rho_L \frac{u_{inj}^2 - \bar{u}^2}{2}} \qquad (1)$$

The above "quadratic formula" gives accurate predictions for the drop size when compared with experimental and correlation results [18]. Earlier work from this laboratory also suggest general applications of this method, and agreements with experimental and correlation data are quite good [20-22]. For liquid sprays in cross flows, it is mostly the gas-phase momentum and kinetic energy that cause the break-up of the liquid column and subsequent atomization. This fact can easily be incorporated in the same formulation as in our previous studies. The geometry is illustrated in Figure 1.

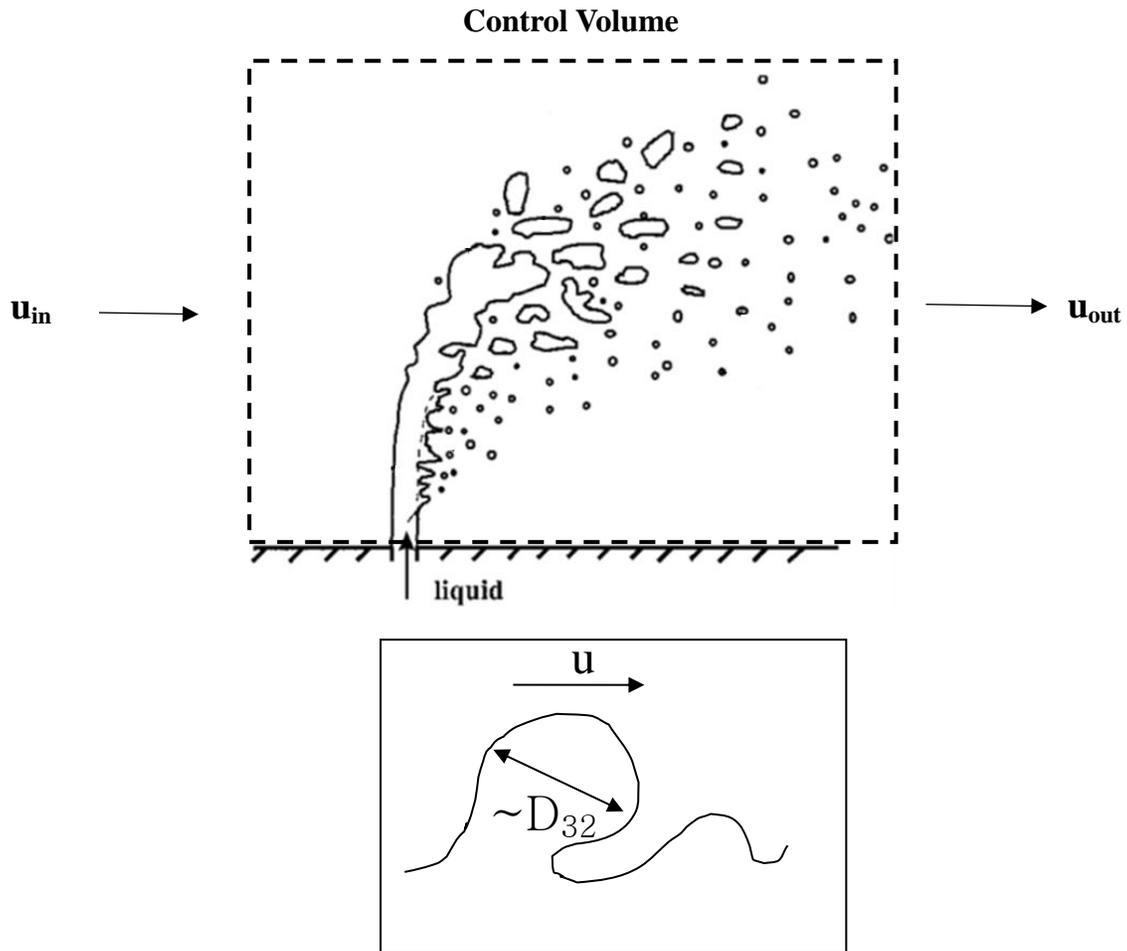

**Figure 1. Basic geometry for the liquid jet atomization in a cross flow (top), and a schematic (bottom) showing the reasoning behind the viscous dissipation term (Eq. 5).**

In Figure 1, liquid is injected vertically into gas flowing left to right, perpendicular to the liquid jet. Here, we focus on perpendicular geometry as shown in Figure 1, although any injection angle relative to the cross flow can potentially be analyzed using trigonometric relations. Relative motion between the liquid and the gas causes the disruption, eventual break-up and atomization of the liquid. Good correlations exist for the liquid jet penetration and trajectories [1-3] for a wide range of conditions. Here, we will resolve the effect of gas-phase velocities, liquid injection and other parameters on the final drop size. A control volume is schematically drawn in Figure 1, which has the cross-sectional area, $A_c$, described later with Eqs. 8 and 9.

As noted earlier, our approach involves direct, robust applications of the conservation laws for mass and energy in integral form [20-22]. Momentum balance can be included to develop relationships between the liquid- and gas-phase velocities, for which iterative solutions are possible [21]. The integral form for the mass and energy balance is:

$$\rho_L u_{inj} A_{inj} = \int_{u=0}^{u_{max}} \int_{D=0}^{D_{max}} np(D,u) \frac{\pi D^3}{6} \rho_L u A \, dD \, du \approx \frac{\pi}{6} n \rho_L u_L A_c \sum_i^N p(D_i) D_i^3 \Delta D_i \qquad (2)$$

$$\rho_L \frac{u_{inj}^3}{2} A_{inj} + \rho_g \frac{u_{in}^3}{2} A_c =$$
$$\frac{\pi}{12} n \rho_L u_L^3 A_c \sum_i^N p(D_i) D_i^3 \Delta D_i + n u_L A_c \pi \sigma \sum_i^N p(D_i) D_i^2 \Delta D_i + \mu_L \left\langle \left(\frac{\partial u}{\partial y}\right)^2 \right\rangle (SprayVolume) + \rho_g \frac{u_{out}^3}{2} A_c$$

(3)

Eq. 2 is a straight-forward mass balance between the injected fluid and the droplet mass flow rate. For the energy balance, we have added the gas-phase kinetic energy entering and leaving the control volume, as parameterized by $u_{in}$ and $u_{out}$, respectively. The input energy is the kinetic energy of the liquid and gas, which is distributed into the droplet kinetic energy, surface tension energy, viscous dissipation, and gas kinetic energy in Eq. 3. As in our previous work, this integral formulation bypasses the complex physics, and relates the input and output terms. From Eq. 2, we can solve for the drop number density, n.

$$n = \frac{\rho_L u_{inj} A_{inj}}{\frac{\pi}{6} \rho_L u_L A_c \sum_i^N p(D_i) D_i^3 \Delta D_i} \qquad (4)$$

Also, it is reasonable to write the viscous dissipation term as [18]:

$$\mu_L \left\langle \left(\frac{\partial u}{\partial y}\right)^2 \right\rangle (SprayVolume) = K\mu_L \left(\frac{u_L}{D_{32}}\right)^2 \qquad (5)$$

The viscous term includes the spray volume, to be consistent with the rest of the terms in Eq. 3. The above phenomenological expression for the viscous dissipation means that on the average the shear stress of the droplet tearing from the liquid surface occurs at the velocity and length scales of the mean liquid velocity and mean drop size, respectively. The liquid is strained by velocity $u_L$, and the length scale over which this strain occurs on the average can be written as $D_{32}$. Schmehl [23] notes that for droplet breakup during secondary atomization the droplet viscous dissipation is exactly $16\pi R_o^3 \left(\frac{\dot{y}}{y}\right)^2$, where $R_o$ is the initial drop radius, y the ellipsoid coordinate and therefore dy/dt the surface velocity. Eq. 5 is mathematically analogous to the expression by Schmehl [23].

Substitution of Eqs. 4 and 5 into Eq. 3 results in the following equation:

$$\left[\frac{\rho_L u_{inj}^2}{2}\left(1-\left(\frac{u_L}{u_{inj}}\right)^2\right)+\frac{u_{in}A_c}{u_{inj}A_{inj}}\frac{\rho_g u_{in}^2}{2}\left(1-\left(\frac{u_{out}}{u_{in}}\right)^3\right)\right]D_{32}^2 - 6\sigma D_{32} - \frac{K\mu_L}{u_{inj}A_{inj}}u_L^2 = 0 \qquad (6)$$

We take the positive (physically meaningful) branch of the quadratic solution of Eq. 6, as follows:

$$D_{32} = \frac{3\sigma + \sqrt{9\sigma^2 + \frac{K\mu_L u_L^2}{u_{inj}A_{inj}}\left[\frac{\rho_L u_{inj}^2}{2}\left(1-\left(\frac{u_L}{u_{inj}}\right)^2\right)+\frac{u_{in}A_c}{u_{inj}A_{inj}}\frac{\rho_g u_{in}^2}{2}\left(1-\left(\frac{u_{out}}{u_{in}}\right)^3\right)\right]}}{\left[\frac{\rho_L u_{inj}^2}{2}\left(1-\left(\frac{u_L}{u_{inj}}\right)^2\right)+\frac{u_{in}A_c}{u_{inj}A_{inj}}\frac{\rho_g u_{in}^2}{2}\left(1-\left(\frac{u_{out}}{u_{in}}\right)^3\right)\right]} \qquad (7)$$

Although Eq. 7 has been obtained from a quadratic equation, we refer to it as a "cubic formula" due to the significance of the gas velocity ratio term, which is cubed. Later (Figure **6**), it is shown that the drop size indeed follows a cubic curve, as a function of the outgoing gas velocity ($u_{out}$). $A_c$ is the cross-sectional area of the control volume enveloping the spray and can be estimated from the liquid jet penetration and width. For example, Wu et al. [2] have performed extensive measurements of the liquid jet geometry, and the following often-cited correlations can be used to estimate $A_c = z_w y_m$, where $x = x_b$ =(column fracture point)=8.06$d_{inj}$. q is the momentum ratio in Eqs. 8 and 9 [2].

$$\frac{z_w}{d_{inj}} = 4.3q^{0.33}\left(\frac{x}{d_{inj}}\right)^{0.33} \qquad (8)$$

$$\frac{y_m}{d_{inj}} = 7.86 q^{0.17} \left(\frac{x}{d_{inj}}\right)^{0.33} \tag{9}$$

Comparing Eqs. 1 (straight) and 7 (cross-flow), we can see that the gas-phase kinetic energy term dominates the cross-flow atomization for $u_{in} \sim 100$ m/s and above. Cross-flow gas velocity is typically high [1-6, 8-10]. Since the kinetic energy flow rate is proportional to the cube of the initial gas velocity, this term can overwhelm the liquid phase kinetic energy for typical momentum ratios. The liquid injection velocities are typically relatively small during cross-flow injections, compared to the gas-phase velocity [1-6, 8-10]. The energy balance (Eq. 6) and its solution (Eq. 7) show that it is the retardation of the gas-phase kinetic energy due to aerodynamic interaction with the liquid-phase that tend to dictate the final drop size. The surface tension energy and viscous dissipation affect the final energy balance as well, and these effects are typically parameterized through the Weber (surface tension effect) and the Reynolds (viscosity) number.

In cross-flow sprays, the liquid column is bent and one can experimentally observe the break-up near the surface extending downstream [1-4]. However, downstream from the location where the tip of the liquid column is completely depleted, one sees a nearly uniform field of droplets. The control volume extends close to the wall in this geometry, but since the flow is parallel to the wall droplet trajectories also become parallel near the wall. Thus, we can define the end of the control volume at a downstream position where the atomization (primary and secondary) has been completed, use the $u_{out}$ at that plane in Eq. 7. There are several examples of good data on $u_{out}$ as the velocities can be measured using PIV or laser/phase Doppler anemometry [1-4, 9, 13-15].

We can see that Eq. 7 includes all the relevant parameters, that can be summarized as the effects of Weber (the surface tension terms), Reynolds (the viscosity terms), injection and gas flow parameters. Eq. 7 is somewhat lengthy but has only three variable parameters to determine $D_{32}$: K, $u_L$ and $u_{out}$. Of these, only K is truly adjustable parameter, as $u_L$ and $u_{out}$ are the mean droplet and outgoing gas velocities, respectively. They will be related to one another through the momentum equation, and approximate values can also be computed iteratively similar to our work in [18, 20, 21]. In that work, momentum balance for the gas- and liquid-phase provided two additional equations to relate $u_{out}$ to $u_{in}$ and $u_L$ to $u_{inj}$. Also, integration to CFD is possible by using the computed velocities to find the initial estimates of the drop size, and iterating until computed velocities and drop size (calculated using Eq. 5) become mutually consistent [20, 21]. In Eqs. 3 and 5, the parameter K prescribes the average viscous dissipation in the control volume, and it has the spray volume term in it [18].

**RESULTS AND DISCUSSION**

Figure 2 shows the effect of the incoming gas velocity, $u_{in}$, on the final drop size using Eq. 7, compared with experimental data of Freitag and Hassa [4] where for low liquid injection velocities (1.3-5.1 m/s) the gas velocity was varied from 50 to 150 m/s. In Figure 2, we can see that there needs to be substantial gas-phase kinetic energy to generate drop size of 100 μm or less. Below $u_{in}$ of approximately 40 m/s, the drop size steeply rises. Dynamically, this has been attributed to the Weber number effect [1-3], where for the Weber number approaching zero, the drop size tends to infinity [1]. Eq. 7 embodies the underlying energy transfer where the kinetic energy ratio of the incoming and outgoing gas is a determinant for the drop size. Eq. 7 simply shows that this dynamical process can be quantitatively assessed based on relative energy between the incoming kinetic and the rest of the energy terms, in particular surface tension energy. Beyond a certain kinetic energy level ($u_{in} \sim 100$ m/s), further decrease in the drop size is gradual and there are diminished returns on increasing the gas-phase velocity. The experimental data show a more gradual decrease of the drop size as a function of the gas velocity than Eq. 7, which may be due to the fact the drop size is a spatial average in the experimental data [4]. A velocity distribution would exist after the gas-liquid momentum exchange, leading also to a range of drop size for a given inlet condition. The experimental data of Freitag and Hassa [4] shows a variation in the SMD (up to 43%) within the measurement plane. Thus, SMD is not uniform, as the velocity field is not uniform. Eq. 5 indeed shows this would be the case if there is a velocity variation. The theoretical results (lines) used in Figure 2 are effectively the specific value of SMD at a given velocity. Since the relationship between the SMD and the velocity is non-linear as shown in Eq. 5, spatial averaging can lead to a some discrepancy. More will be said later on this relationship between the drop size and final gas velocity ($u_{out}$) distribution. The experimental data show a weak dependence on the injector diameter, where the drop size increases with the injector diameter [4].

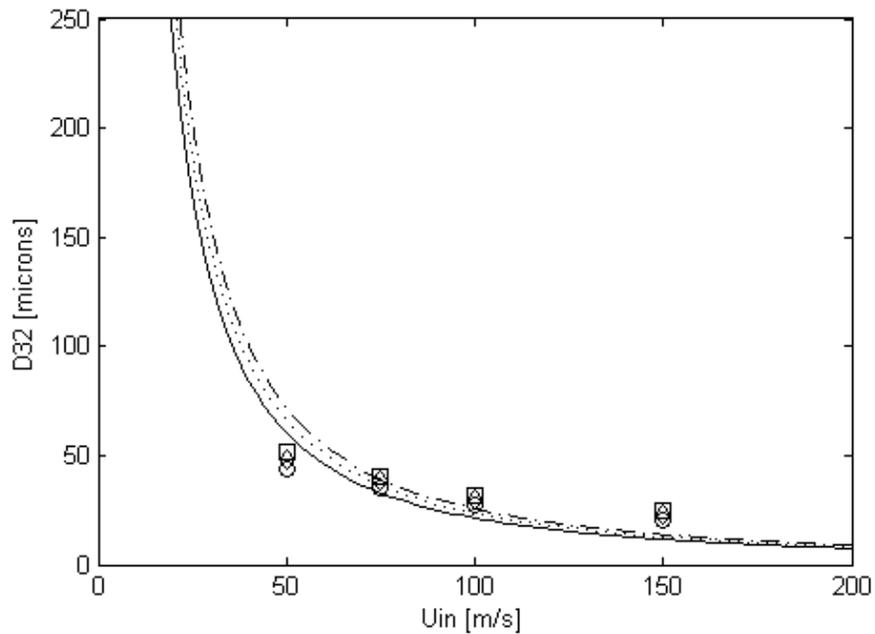

**Figure 2. Effect of incoming gas velocity on the SMD. Three different injector diameters: 0.3 mm (circle, solid line), 0.5 (diamond, dotted line), and 0.7 (square, dashed line). The experimental data [4] are plotted as symbols.**

Figure 3 is a similar comparison showing the effects of gas density or the gas pressure. In Freitag and Hassa [4], the air pressure was varied from 0.2 to 0.8 MPa. The pressure effect appears as the gas density effect in Eq. 7, as the gas density will increase proportionally with the pressure. Since the gas-phase kinetic energy flow rate is linear with the density (as opposed to cubic dependence on the velocity), the decrease in the drop size with increasing density is relatively small in Figure 3. This is due to the cubic term for $u_{out}/u_{in}$ in Eq. 7. For large $u_{in}$, the kinetic energy contribution to the drop formation rapidly approaches its maximum.

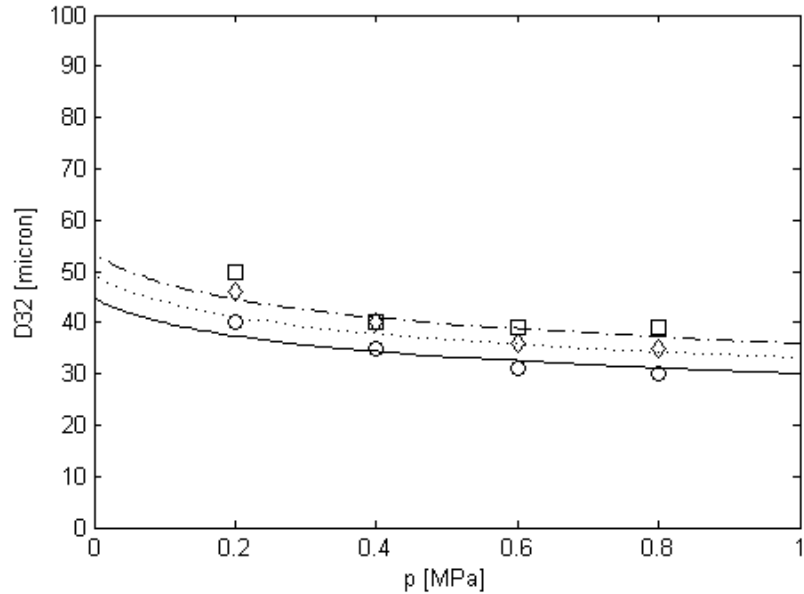

**Figure 3. Effect of gas pressure (density) on the SMD. Three different injector diameters: 0.3 mm (circle, solid line), 0.5 (diamond, dotted line), and 0.7 (square, dashed line). The experimental data [4] are plotted as symbols.**

Other authors present the data in different ways, and drop size dependence on relevant injection parameters can still be retrieved from some of those data sets. For example, Shafaee et al. [5] also present the SMD measured at a fixed location, as a function of the gas velocity for various injector diameters. These results are plotted in Figure 4, where injector diameters were 0.8, 1.2, and 1.6 mm. Although the injector diameters were varied, the mass flow rate was fixed. Thus, the injector diameter essentially varied the momentum ratio and also the spray volume. The parameter, K, in Eq. 7 for the viscous dissipation term contains the spray volume term. Any changes in the spray volume should be taken into account where the spray volume would increase with the injector diameter. K increases with the injector diameter (spray volume), and they are adjusted to provide an agreement at a given $u_{in}$. K is then held fixed for each injector diameter.  K were set at 0.9, 6.7, 18 and 33, respectively, for injector diameters of 0.8, 1.2, 1.6 and 1.8 mm.

The drop size change, as a function of the gas velocity for various injector diameters (momentum ratios), is again quite well tracked by Eq. 7, as shown in Figure 4.  Figure 4 also shows that the change in SMD is gradual relative to the gas velocity beyond a certain gas velocity range (about $u_{in}$ = 100 m/s in Figure 4).  Similar observations have been made where SMD change was relatively small in comparison to large increases in the gas velocities [4-6].  The agreement between the data and theory is quite good for $d_{inj}$=0.8 and 1.2 mm in Figure 4.  For larger injector diameters (1.6 and 1.8 mm), the change in SMD with increasing $u_{in}$ is small in comparison to the current results.  For larger injector diameters, the spray volume is larger, and therefore the potential for spatial averaging bias (mentioned earlier) is larger.  Larger drop sizes also mean possible presence of non-spherical or even ligaments, which will not appear in drop size measurements based on sphericity of droplets, thus biasing down the measured SMD. Nonetheless, the trend toward increase in drop size relative to $u_{in}$ and $d_{inj}$ are well reproduced by the theoretical lines.

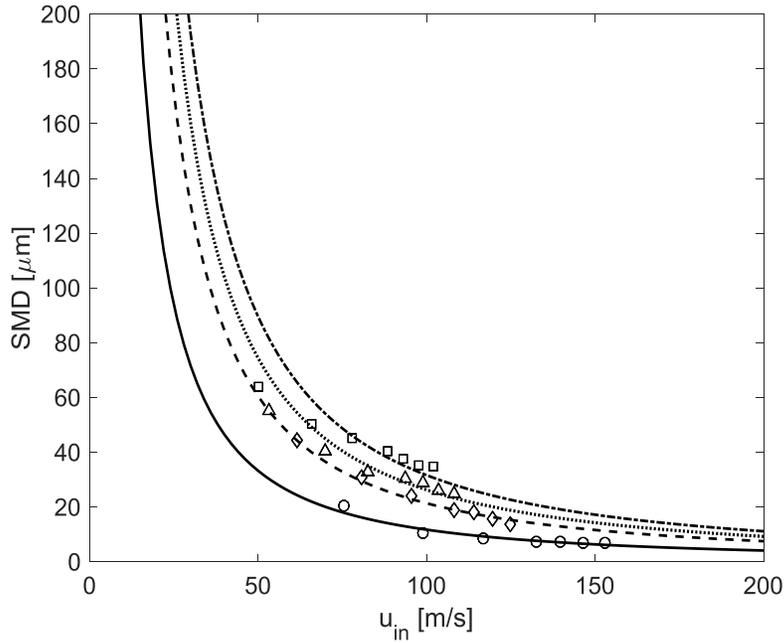

**Figure 4.** Comparison of the measured and calculated SMD. Four injector diameters were used: $d_{inj}$ = 0.8 mm (○, solid line), 1.2 (◊, dashed), 1.6 (Δ, dotted) and 1.8 (□, dash-dot).

Kihm et al. [6] presents a drop size correlation in terms of Reynolds and Weber numbers along the spray arc length (x and y in their notations). We usually prefer to apply Eq. 7 to a fixed position far downstream, as the drop size data typically exhibit decreasing trend with increasing distance from the injectors. However, the asymptotic location (the location where the drop size no longer changes) is not easily reached, either dynamically or experimentally. Also, in an attempt to account for all of the relevant parameters, such as gas velocity, injection velocity, densities, surface tension and viscosities, the range of gas velocity that was varied in the experiments [6] was not significant enough to produce a large change in the drop size, as shown in Figure 5. $y/d_{inj}$ locations varied from 35, 40 to 45 in Kihm et al. [6]. In spite of the fact that the measurement locations are varied, we can argue that the $u_{out}/u_{in}$ to be input into Eq. 7 should decrease as the liquid and gas phases exchange momentum further along the liquid trajectory, resulting in liquid phase gaining momentum and the gas phase losing momentum.   Fig. 5 shows a comparison of the drop size with that calculated using Eq. 7, where $u_{out}$ of 0.57, 0.75, and $0.85u_{in}$ were used for $y/d_{inj}$ locations of 35, 40 and 45, respectively.   K was fixed at 2.55.   Although this required some optimization of the unknown $u_{out}/u_{in}$ ratio, the trend in the drop size as a function of gas velocity is captured quite closely at various locations in the spray.   As in our previous work [20] momentum equations for the gas and liquid phase can be used to estimate $u_{out}/u_{in}$, or computational simulations can be augmented with Eq. 7 to iteratively determine the drop size and $u_{out}/u_{in}$.

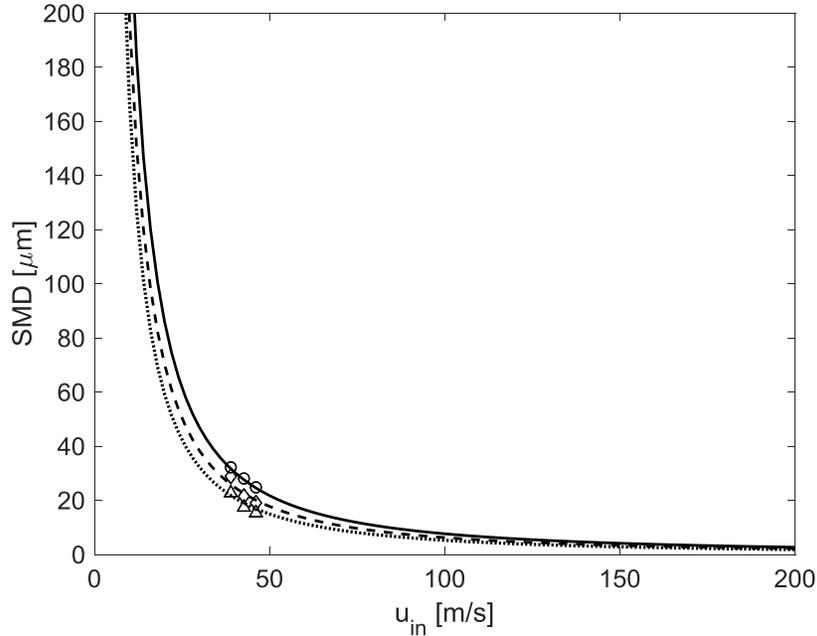

**Figure 5.   Comparison of the measured and calculated SMD at various locations.   $y/d_{inj}$ = 35, solid line (○); $y/d_{inj}$ = 40, dashed (◊), and $y/d_{inj}$ = 45, dotted (Δ).**

Above results, and Eq. 7, show that crossflow atomization under typical injection conditions is primarily determined by the ratio of the incoming to outgoing gas velocity.   The reason is evident from the energy balance: if more of the kinetic energy of the gas that goes into the surface tension energy, then smaller droplets will be formed. In other words, the larger the difference (or the ratio) between the outgoing and incoming gas velocities, the drop size will become smaller. We saw this effect of incoming gas velocity in Figure 2. Figure 6 is a representation of the effect of $u_{out}/u_{in}$ on the drop size, and also can be used as a direct relationship between the velocities and drop size. Eq. 5 points to a cubic increase in the drop size with $u_{out}/u_{in}$. When $u_{out} \rightarrow u_{in}$, no gas-phase kinetic energy has gone into the atomization process, and the drop size is infinite. Dynamically, the aerodynamic interaction between the liquid and gas phase will result in momentum exchange through the drag force, with the gas-phase momentum going into liquid momentum in the streamwise (for the gas-flow) direction. This will also be represented as kinetic energy loss for the gas phase, along with the requisite surface tension energy when droplet curvatures are generated according to Eq. 7. Thus, the larger the energy loss, or equivalently the lower the outgoing gas velocity relative to a fixed $u_{in}$, smaller drop size will result. It is interesting to note that below $u_{out}/u_{in} \sim 0.8$, the drop size change is relatively gradual, with drastic changes occurring for $u_{out}/u_{in}$ from about 0.85 to 1, due to the cubic shape of the curve. A similar behavior has been observed for secondary atomization where the kinetic energy difference between the incoming and exiting gas causes the parent droplet to break up into a group of small droplets [22].

In Figure 1, the control volume is drawn to envelop the entire spray, where the average velocities at the inlet and exit are used in Eq. 7. However, as in our previous study [18], localized control volumes can be drawn (e.g. horizontal strips) in Figure 1. Then, local velocities can be used in Eq. 7 to find spatial distribution of $D_{32}$. For experimental data, this would require spatially-resolved velocity data. For computational simulations, local velocities can be used for estimation of the drop size at specific locations. Some examples of this application to computational simulations have been shown in our previous work [18].

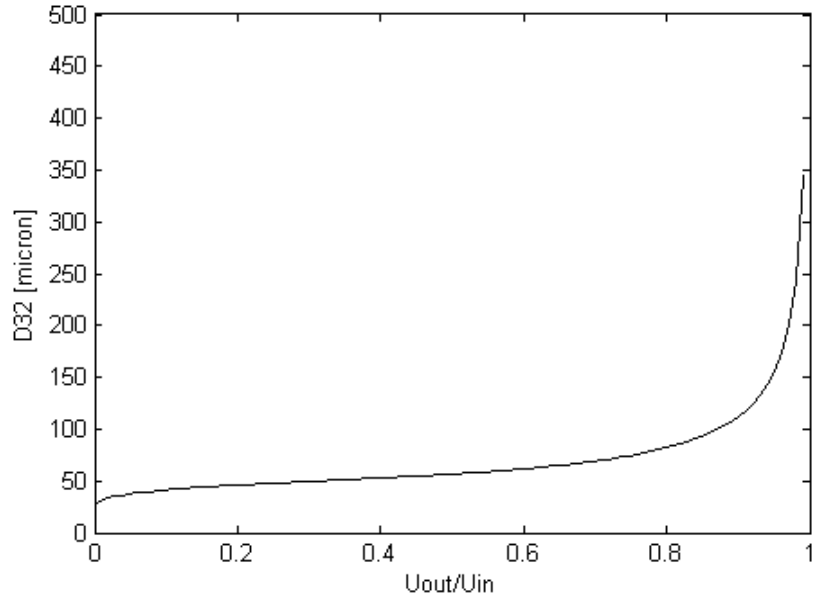

**Figure 6.  Effect of the gas-phase velocity ratio on the SMD.**

Figure 6 also hints at a self-consistent method for embedding the cubic formula (Eq. 7) in computations of sprays in cross flows.   For example, data on gas velocity distribution at the exit plane can be converted to the drop size distribution through Eq. 7.   This process is illustrated in Figure 7, where starting from some gas velocity distribution (Gaussian or reversed log-normal in Figure 7(a)) we can generate the drop size distributions using Eq. 7 as shown in Figure 7(b).   Eq. 7 can therefore be used as the drop size-velocity cross-correlator.   In Figure 7(b), the reversed log-normal distribution leads to a drop size distribution with a wide spread, while the Gaussian results in a distribution sharply peaked at a small drop diameter.

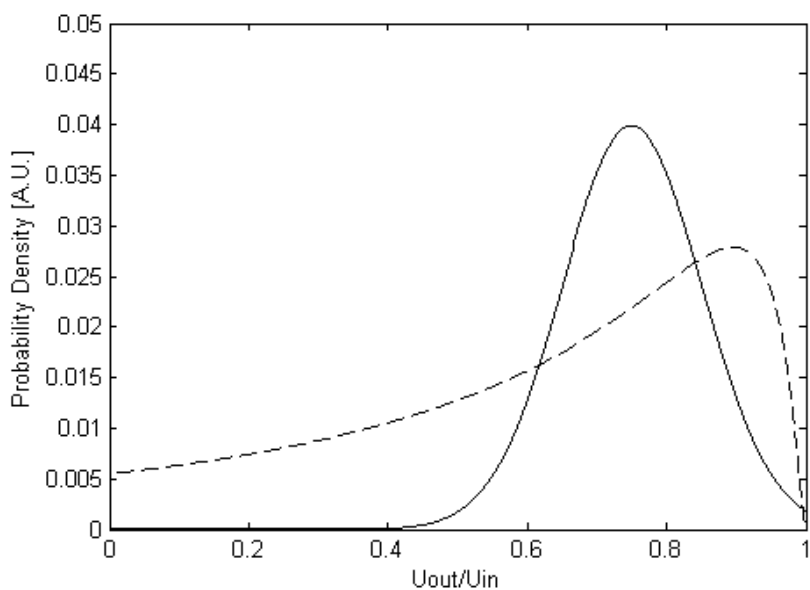

**Figure 7(a).** Examples of probability density function for the velocity ratio ($U_{out}/U_{in}$).

-------- (Gausian);   - - - - (reversed log-normal)

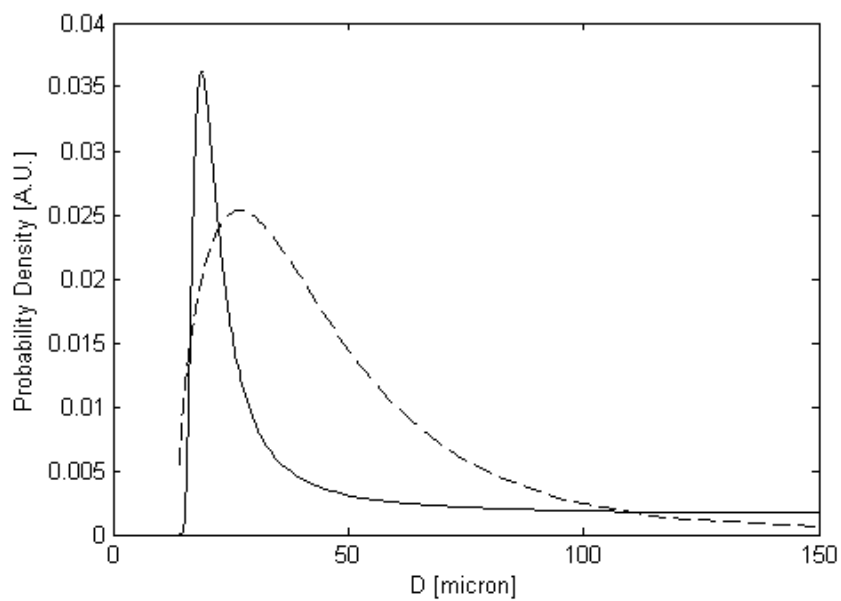

**Figure 7(b).** Resulting drop size distribution using Eq. 7.

-------- (from Gausian);   - - - - (from reversed log-normal)

**CONCLUSIONS**

We have used a theoretical framework based on the integral form of the conservation equations, to find a cubic formula for the drop size as a function of spray parameters, for liquid jet atomization in cross flows. The energy equation leads to the drop size varying as a cubic function of the gas-phase velocity ratio, which agrees with observed data as well as intuition. This solution (Eq. 7) can be used as a sub-routine to prescribe the drop size in computation fluid dynamics of liquid jets in cross flows. An initial estimated drop size or distribution can be specified along the liquid jet surface (e.g. Eqs. 8 and 9), and the drop velocities can be tracked while exchanging momentum with the gas-phase through the drag force. The resultant liquid and gas velocities can be used to check whether the initial drop size was close to the value as given by Eq. 7. This process can be iterated until a satisfactory match between the initial and final drop size is achieved.

The current approach as in our past work [18, 20-22] avoids any ad-hoc modeling or assumptions, and uses the conservation laws in integral (enveloping control volume) sense. Although some discrepancies exist, due to various reasons discussed above, refinements in estimating the viscous dissipation term and inputting the complete gas- and liquid-phase velocities can bring the results closer together. Some experimental correlations exist [e.g. 1, 11, 12]; however, as noted at the outset some of these correlations only provide the primary break-up drop size. Also, due to a large number of parameters, it is difficult to run experiments to come up with a general correlation that incorporates and validates the effects of all the relevant variables. Current analysis and results can synthesize the existing data, and can be extended to other fluids and spray conditions, since they are based on fundamentally correct physics of spray atomization (conservation laws). Thus, the current approach represents a viable method for dealing with this complex spray atomization process, and bears both practical and fundamental significance.